\documentclass[copyright,creativecommons]{eptcs}
\providecommand{\event}{ICLP 2020} % Name of the event you are submitting to
\usepackage{breakurl}             % Not needed if you use pdflatex only.
\usepackage{underscore}           % Only needed if you use pdflatex.
\usepackage{mathptmx}
\usepackage{color}
\usepackage{url}
\usepackage{xspace}
\usepackage{graphicx}
\usepackage{lipsum}
\usepackage{wrapfig} 
\usepackage{multirow}
\usepackage{booktabs}
\usepackage{comment}
\usepackage{subcaption}
\newtheorem{example}{Example}

\newcommand\bcmdtab{\noindent\bgroup\tabcolsep=0pt%
  \begin{tabular}{@{}p{10pc}@{}p{20pc}@{}}}
\newcommand\ecmdtab{\end{tabular}\egroup}

\newcommand\Kscr{\ensuremath{\mathcal{K}}}

\newcommand\ie{\emph{i.e.}\xspace}
\newcommand\eg{\emph{e.g.}\xspace}

\makeatletter
\newcommand\CO[1]{%
  \@tempdima=\linewidth%
  \advance\@tempdima by -2\fboxsep%
  \advance\@tempdima by -2\fboxrule%
  \leavevmode\par\noindent%
  \fbox{\parbox{\the\@tempdima}{%
    \small\normalfont\sffamily #1}}%
  \smallskip\par}
\makeatother

\title{Less Manual Work for Safety Engineers: Towards an Automated Safety Reasoning with Safety Patterns}
\author{Yuri Gil Dantas \qquad\qquad Antoaneta Kondeva \qquad\qquad Vivek Nigam
\institute{fortiss GmbH\\
Research Institute of the Free State of Bavaria\\
Guerickestra{\ss}e 25\\80805 M{\"u}nchen, Germany}
\email{dantas@fortiss.org \qquad\quad kondeva@fortiss.org \quad\qquad nigam@fortiss.org}
}
\def\titlerunning{Towards an Automated Safety Reasoning with Safety Patterns}
\def\authorrunning{Y. G. Dantas, A. Kondeva \& V. Nigam}
\begin{document}
\maketitle
\sloppy

\begin{abstract}
The development of safety-critical systems requires the control of hazards that can potentially cause harm. 
To this end, safety engineers rely during the development phase on architectural solutions, called safety patterns, such as safety monitors, voters, and watchdogs. 
The goal of these patterns is to control (identified) faults that can trigger hazards.
Safety patterns can control such faults by e.g., increasing the redundancy of the system.
Currently, the reasoning of which pattern to use at which part of the target system to control which hazard is documented mostly in textual form or by means of models, such as GSN-models, with limited support for automation.
%Currently, the reasoning of which pattern to use at which part of the target system to control which hazard is documented mostly in textual form or by means of models, such as GSN-models.
%This provides limited support for automation.
This paper proposes the use of logic programming engines for the automated reasoning about system safety. 
We propose a domain-specific language for embedded system safety and specify as disjunctive logic programs reasoning principles used by safety engineers to deploy safety patterns, \eg, when to use safety monitors, or watchdogs.
Our machinery enables two types of automated safety reasoning: (1) identification of which hazards can be controlled and which ones cannot be controlled by the existing safety patterns; 
and (2) automated recommendation of which patterns could be used at which place of the system to control potential hazards. 
Finally, we apply our machinery to two examples taken from the automotive domain: an adaptive cruise control system and a battery management system.
\end{abstract}

\section{Introduction}
\label{sec:intro}
%!TEX root = iclp20.tex

The development of safety-critical systems, such as vehicles, aircraft and medical devices aims to achieve two goals: (1) to develop systems that cannot cause any harm, and (2) to convince regulatory bodies about the safeness of the system by demonstrating compliance to safety standards~\cite{iso26262,arp4754a}.

To achieve the first goal, safety engineers perform safety analysis to ensure that systems cannot cause any harm. 
For example, \emph{Hazard Analysis}~\cite{iso26262,ar4761} identifies the main hazards that shall be controlled.
Other safety techniques, \eg, FTA~\cite{ar4761}, STPA~\cite{stpa}, FMEA~\cite{ar4761}, HAZOP~\cite{hazop}, break down the identified main hazards into component hazards (a.k.a component failures), \ie, faults that can trigger main hazards. 
Safety engineers commonly use \emph{safety architectural patterns}~\cite{preschern13plop,martin,valdivia14dasc} to control the identified component hazards (or hazards for short) thus controlling the main hazards.
To achieve the second goal, safety engineers shall develop a safety case ~\cite{iso26262,defenceUK} for the system under development. 
The purpose of the safety case is to both (a) ensure that all hazards have been analyzed and (b) answer why a safety pattern has been deployed at a particular component to control which hazard.

Safety cases are often documented in textual form, or by models \eg, the Goal Structure Notation (GSN)~\cite{gsn11standard}.
These models, however, have limited support for automated reasoning~\cite{kondeva19wosocer}.  
It is not possible to automatically check whether safety arguments used in a safety case are correct, \ie, check whether all hazards have been controlled by, \eg, safety patterns. This is because the \emph{safety reasoning} used to support system safety is implicitly written textually thus lacking the precise semantics to enable automation~\cite{nigam18safsec}.
As a result, correctness checks are performed manually, possibly leading to human errors.

Our vision is to build an incremental development process for system safety and security assurance cases using automated methods that incorporate safety and security reasoning principles. 
This paper is the first step towards achieving this vision.
We provide safety reasoning principles with safety patterns used during the definition of system architecture for embedded systems.
We specify these principles using logic and logic programming as they are suitable frameworks for the specification of reasoning principles as knowledge bases and using them for automated reasoning~\cite{baral.book}. 

Our main contributions are threefold:

\begin{itemize}
	\item \textbf{Domain-Specific Language (DSL):} We propose a DSL for safety reasoning with safety patterns. Our DSL includes (1) architectural elements, both functional components and logical communication channels; (2) safety hazards including guidewords used in typical analysis, \eg, erroneous or loss of function; (3) a number of safety patterns including n-version programming, safety monitors, and watchdogs;
	
	\item \textbf{Reasoning Principles:} We specify key reasoning principles for determining when a hazard can be controlled or not, including reasoning principles used to decide when a safety pattern can be used to control a hazard. These reasoning principles are specified as Disjunctive Logic Programs~\cite{eiter97tds} based on the DSL proposed;
	
	\item \textbf{Automation:} We illustrate the increased automation enabled by the specified reasoning principles using the logic programming engine DLV~\cite{leone06tcl}. Our machinery enables two types of automated reasoning: \emph{(1) Controllability:} which hazards can be controlled by the given deployed safety patterns and which hazards cannot be controlled. \emph{(2) Safety Pattern Recommendation:} which safety patterns can be used and where exactly they should be deployed to control hazards that have not yet been controlled.
\end{itemize}

We validate our machinery\footnote{All machinery needed to reproduce our results are publicly available: \url{https://github.com/ygdantas/safpat}} with two examples of safety-critical embedded systems taken from the automotive domain. The first example is an \emph{Adaptive Cruise Control} system installed in a vehicle to adapt its speed in an automated fashion without crashing into objects in front and at the same time trying to maintain a given speed. The second example is a \emph{Battery Management System}~\cite{martin} responsible for ensuring that a vehicle battery is charged without risking it to explode by, \eg, overheating. Our machinery infers a number of possible solutions involving different safety patterns that can be used to control identified hazards. 

\begin{comment}
The remainder of the paper is structured as follows. 
Section~\ref{sec:motivation} describes two motivating examples. 
Section~\ref{sec:prelim} briefly describes basic notations of safety patterns, answer-set programming and disjuctive logic programs.
Sections~\ref{sec:dsl},~\ref{sec:automation}, and~\ref{sec:recommendation} describe our main contributions.
Section~\ref{sec:examples} illustrates the types of results that can be delivered by our machinery.
%Section~\ref{sec:examples} validates our machinery with the two motivating examples described in Section~\ref{sec:motivation}.
%Finally, in Sections~\ref{sec:related} and~\ref{sec:conc}, we conclude by discussing related and future work.
%Section~\ref{sec:related} discusses the related work, and the paper is concluded in Section~\ref{sec:conc}.}
After a discussion of the related work in Section~\ref{sec:related}, the paper is concluded in Section~\ref{sec:conc}.
\end{comment}

\section{Motivating Examples}
\label{sec:motivation}
%!TEX root = iclp20.tex
\newcommand\acc{\ensuremath{\mathsf{ACC}}\xspace}
\newcommand\accm{\ensuremath{\mathsf{ACCM}}\xspace}
\newcommand\ds{\ensuremath{\mathsf{DS}}\xspace}
\newcommand\vs{\ensuremath{\mathsf{VS}}\xspace}
\newcommand\bs{\ensuremath{\mathsf{BS}}\xspace}
\newcommand\ps{\ensuremath{\mathsf{PS}}\xspace}

\newcommand\accerr{\textbf{H1$_{acc}$}\xspace}
\newcommand\accloss{\textbf{H2$_{acc}$}\xspace}
\newcommand\derr{\textbf{H1.1$_{acc}$}\xspace}
\newcommand\verr{\textbf{H1.2$_{acc}$}\xspace}
\newcommand\accmerr{\textbf{H1.3$_{acc}$}\xspace}

\newcommand\bms{\ensuremath{\mathsf{BMS}}\xspace}
\newcommand\fw{\ensuremath{\mathsf{FW}}\xspace}
\newcommand\bat{\ensuremath{\mathsf{BAT}}\xspace}
\newcommand\ci{\ensuremath{\mathsf{CI}}\xspace}

\newcommand\mainhzbms{\textbf{H0$_{bms}$}\xspace}
\newcommand\bmserr{\textbf{H1.1$_{bms}$}\xspace}
\newcommand\cierr{\textbf{H1$_{bms}$}\xspace}
\newcommand\canerr{\textbf{H1.2$_{bms}$}\xspace}
\newcommand\fwerr{\textbf{H1.3$_{bms}$}\xspace}

\newcommand\can{\ensuremath{\mathsf{CAN}}}

This section describes two examples from the automotive domain. 
We refer to these examples as Adaptive Cruise Control system (\acc) and Battery Management System (\bms). 
We use the \acc as a running example throughout the paper. 
We get back to the \bms example in Section~\ref{sec:examples}.

\paragraph{Adaptive Cruise Control (ACC).}
Consider as a motivating example, a simplified \acc responsible for maintaining safe distance to objects in front of its vehicle. 
The \acc is a critical system as harm, \eg, accidents, may occur if the \acc is faulty.
%The \acc is safety-critical as harm, \eg, accidents, may occur if the \acc is faulty.

\begin{figure}[t]
  \centering
  \includegraphics[width=0.6\textwidth]{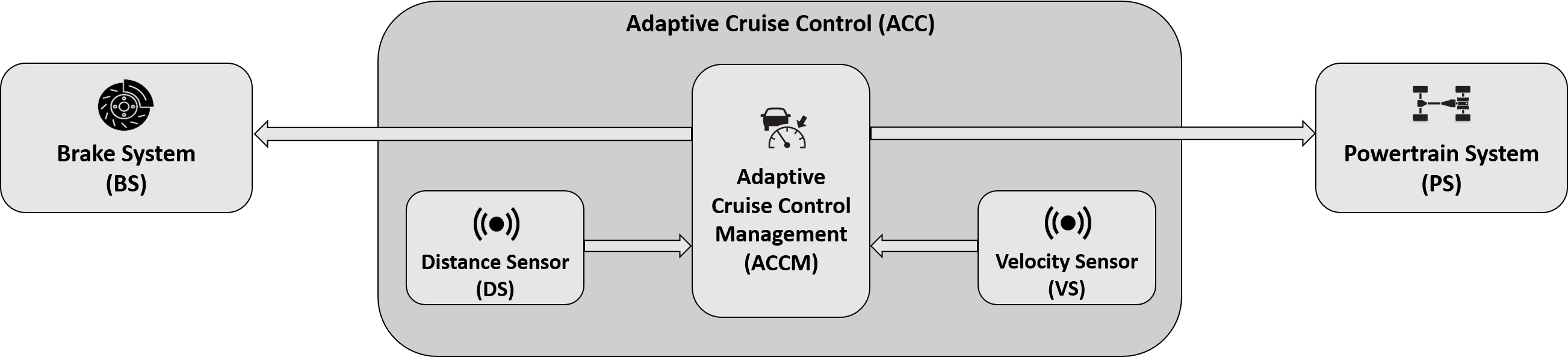}
  \caption{Adaptive Cruise Control (\acc) Functional Architecture}
  \label{fig:acc-func}
\end{figure}

Figure~\ref{fig:acc-func} depicts the main functions composing the \acc.
% These functions are taken from the AutoFOCUS implementation of small scale rovers~\cite{af3-rovers}. 
\acc uses information from two sensing functions: (1) distance sensor function (\ds) that computes the distance to objects immediately in front; (2) velocity sensor function (\vs) that computes the vehicle's current speed. 
The ACC Management function (\accm) computes (adequate) acceleration and braking values for the vehicle which are sent to the power-train control (\ps) and brake control functions (\bs), respectively. 
Notice that \ps and \bs are not part of the \acc but interact with the \acc.

%\paragraph{Hazards.}
To address the safety of the \acc, safety analysis are carried out, such as Hazard Analysis, to determine main hazards. The main hazard is:
\newcommand\mainhz{\textbf{H0$_{acc}$}\xspace}

\begin{center}
  \mainhz: The vehicle does not maintain a safe distance to any object in front.
\end{center}

%When related to the \acc, 
We identify two hazards, $\accerr$ and \accloss, that may lead to \mainhz. 
The words loss and erroneous are used by safety engineers to describe hazards: \emph{loss} is used when a hazard is triggered whenever a function is not working, and \emph{erroneous} when a function is working but not correctly. 
%The former is used when a hazard is triggered whenever a function is not working, and the latter when a function is working by not correctly. 

\begin{itemize}
  \item \textbf{\accerr -- Erroneous ACC:} \acc computes incorrect acceleration or braking values;
  \item \textbf{\accloss -- Loss of ACC:} \acc is not functioning.
\end{itemize}

%These hazards are subsequently further broken down to identify which sub-functions can trigger them using, \eg, Fault Tree Analysis. The following sub-hazards can lead to \textbf{H1}:
These hazards are subsequently further broken down to identify which sub-functions can trigger them using, \eg, Fault Tree Analysis. The following hazards may lead to \textbf{H1}:

\begin{itemize}
  \item \textbf{\derr- Erroneous DS:} The \ds computes an incorrect distance to the car in front;
  
  \item \textbf{\verr- Erroneous VS:} The \vs computes an incorrect velocity;

  \item \textbf{\accmerr- Erroneous ACCM:} The \accm computes wrong acceleration or braking values.
\end{itemize}

\paragraph{Battery Management System (BMS).} We consider a simplified \bms responsible for controlling a rechargeable electric car battery~\cite{martin}. 
The \bms is a critical system as harm, \eg, battery explosions, may occur if it does not compute the charging state of the battery correctly. 

\begin{figure}[t]
  \centering
  \includegraphics[width=0.4\textwidth]{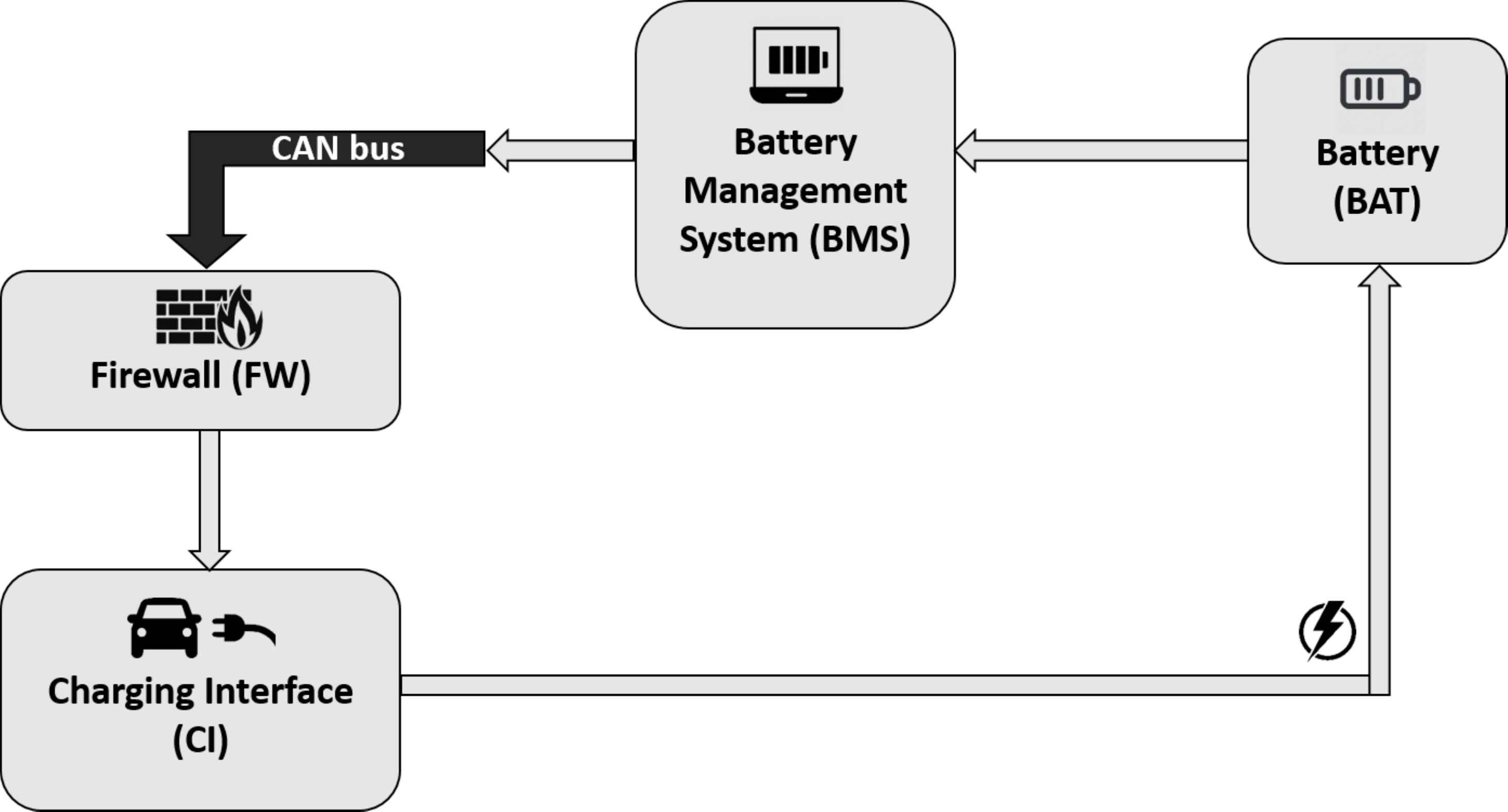}
  \caption{Battery Management System (\bms) Functional Architecture}
  \label{fig:bms-func}
\end{figure}

Figure~\ref{fig:bms-func} depicts the main functions composing the \bms.
%The architecture of the system is depicted in Figure~\ref{fig:bms-func}.
The charging interface (\ci) represents the interface at the charging car station.
This interface is triggered while recharging the battery (\bat) of the car.
%\bat is the actual battery of the electric car, while \ci represents the interface at the charging car station.
\bms receives relevant information (\eg, voltage and temperature values) from \bat so that it can compute the charging state of \bat.
%\bat sends relevant information (\eg, voltage and temperature values) to \bms so that it can control the state of \bat.
Depending on the state of \bat, \bms sends signals of activation or deactivation of the external changer to \ci.
%Depending of the state of \bat, \bms sends signals of activation or deactivation of external changer to \ci.
These signals are sent though a \can\ bus.
\ci is considered the only function accessible by external users (e.g., drivers). 
To avoid that an intruder can access the \can\ bus through \ci, a firewall (\fw) is placed between \bms and \ci.\footnote{We refer the reader to~\cite{martin} for more insights on why adding a \fw between \bms and \ci makes the system more secure.} 
This decision, however, comes at a safety impact, as mentioned below.
%\red{Hence, for security reasons, a firewall (\fw) is placed between \bms and \ci.}
The main hazard considered here is:

\begin{center}
  \mainhzbms: The \bat is overcharged leading to its explosion.
\end{center}

%We identified one erroneous hazard \cierr that may lead to \mainhzbms: The \ci sends charging signals when \bat is fully charged. 
We identify one erroneous hazard \cierr that may lead to \mainhzbms. 

\begin{itemize}
  \item \textbf{\cierr -- Erroneous CI:} The \ci sends charging signals when \bat is fully charged. 
\end{itemize}

The following three hazards may lead to \cierr.
We use the word \emph{omission} as a specialization of the erroneous behavior whenever the corresponding function does not provide an output when such an output is expected, \eg, not outputting a fail-safe signal.

\begin{itemize}
  \item \textbf{\bmserr -- Erroneous BMS:} The \bms sends wrong signals to \ci;
  \item \textbf{\canerr -- Erroneous CAN:} The \can\ bus sends wrong signals to \ci;
  \item \textbf{\fwerr -- Omission FW:} The \fw incorrectly blocks signals from \bms.
\end{itemize}

Hazards are also associated with  severity class denoting the level of harm it can cause. Severity classes range over \emph{no effect}, \emph{minor}, \emph{major}, \emph{fatal}, and \emph{catastrophic}. 
The hazards described in this section are classified as \emph{catastrophic}, which means that they shall be strongly controlled.

\section{Preliminaries}
\label{sec:prelim}
%!TEX root = iclp20.tex
%This section reviews basic notions about safety patterns, ASP and disjuctive logic programs.
%This section reviews some basic notions on safety patterns, ASP and disjuctive logic programs.

\paragraph{Safety Architectural Patterns.}
\newcommand\tmoon{\ensuremath{\mathsf{TMR}}\xspace}
\newcommand\dmoon{\ensuremath{\mathsf{HDR}}\xspace}
\newcommand\safmon{\ensuremath{\mathsf{SafMon}}\xspace}
\newcommand\watch{\ensuremath{\mathsf{WD}}\xspace}
\newcommand\sanity{\ensuremath{\mathsf{San}}\xspace}
\newcommand\nprog{\ensuremath{\mathsf{NProg}}\xspace}

In the architectural level, a number of safety patterns are typically used for embedded system safety~\cite{valdivia14dasc,preschern13plop}. Examples of such patterns are Heterogeneous Duplex Redundancy ($\dmoon$), Triple Modular Redundancy ($\tmoon$), N-Version Programming ($\nprog$), Safety Monitors ($\safmon$), and Watchdog ($\watch$).

%Each pattern is used to achieve control over some type of hazards provided some conditions are satisfied. 
The goal of these patterns is to control some type of hazards provided some conditions are satisfied.
$\watch$s are used to detect when there is loss of function, thus controlling hazards associated with a \emph{loss} of function.
$\safmon$s are used to check whether a function is computing correctly, thus controlling hazards associated with \emph{erroneous} functions. 
%Other safety patterns, such as 
\dmoon and \tmoon are used to control hazards by increasing the redundancy of existing hardware, thus reducing the overall fault rate. 
They can also be used to increase the redundancy of paths in the system in case messages are lost or incorrectly computed. 
$\nprog$s are used control hazards associated with possibly \emph{erroneous} software functions by increasing the redundancy of such functions.

\paragraph{Answer-Set Programming and Disjunctive Logic Programs.}
We assume that the reader is familiar with Answer-Set Programming (ASP) and provide only a brief overview here. Let $\Kscr$ be a set of propositional variables. 
A \emph{default literal} is an atomic formula preceded by \emph{not}. 
A propositional variable and a default literal are both \emph{literals}. A rule $r$ is an ordered pair $Head(r) \leftarrow Body(r)$, where $Head(r) = \ell$ is a literal and $Body(r) = \{\ell_1, \ldots, \ell_n\}$ is a set of literals. 
Such a rule is written as $\ell \leftarrow \ell_1, \ldots, \ell_n$. An \emph{Answer-Set Program} (LP) is a set of rules. 
An interpretation $M$ is an \emph{answer set} of a LP $P$ if $M' = least(P \cup \{not\_A \mid A \notin M\})$ and $M'= M \cup \{not\_A \mid A \notin M\}$, where least is the least model of the \emph{definite logic program} obtained from the program $P$ by replacing all occurrences of $not~A$ by a new atomic formula $not\_A$.

% This assumes that the vocabulary $\Kscr$ contains the predicates and constants described in Section~ \dsl.

The interpretation of the default negation $not$ assumes a \emph{closed-world} assumption. That is, we assume to be true only the facts that are explicitly supported by a rule. For example, the following program $P$ with three rules has two answer-sets $\{a,c\}$ and $\{b\}$:
\[
  a \leftarrow not~b \qquad b \leftarrow not~a \qquad c \leftarrow a
\]
DLV is an engine implementing disjunctive logic programs~\cite{eiter97tds} based on ASP semantics~\cite{gelfond90iclp}. 
In particular, a rule may have disjunction in its head, \eg, $a_1 \lor \cdots \lor a_m \leftarrow \ell_1, \ldots, \ell_n$, where $a_i$ for $0 \leq i \leq m$ are atomic formulas. 
For example, consider the program $P_1$ with the two clauses $a \lor b$ and $c \leftarrow a$. It has the same two answer-sets as the program $P$. 
If a rule's  head is empty, \ie, $m = 0$, then it is a constraint. 
For example, if we add the clause $\leftarrow b$ to $P_1$, then the resulting program has only one answer-set $\{a,c\}$.

In the remainder of this paper, we use the DLV notation writing \texttt{:-} for $\leftarrow$ and \texttt{v} for $\lor$. For example, the program $P_1$ is written as \texttt{a v b} and \texttt{c :- a}. 

% \section{Answer Set Programming}
% \red{Vivek}

%\section{Basic Domain-Specific Language: Functional, Hardware and Safety Patterns}
\section{Basic DSL: Functional, Hardware and Safety Patterns}
\label{sec:dsl}

%!TEX root = iclp20.tex
\newcommand\dsl{\ensuremath{\mathsf{SafPat}}}
\newcommand\cp{\ensuremath{\mathsf{cp}}}
\newcommand\subcp{\ensuremath{\mathsf{subcp}}}
\newcommand\ch{\ensuremath{\mathsf{ch}}}
\newcommand\info{\ensuremath{\mathsf{if}}}
\newcommand\hw{\ensuremath{\mathsf{hw}}}
\newcommand\sw{\ensuremath{\mathsf{sw}}}
\newcommand\dep{\ensuremath{\mathsf{dep}}}
\newcommand\id{\ensuremath{\mathsf{id}}}
\newcommand\ecu{\ensuremath{\mathsf{ecu}}}
\newcommand\Dep{\ensuremath{\mathsf{dep}}}
\newcommand\hz{\ensuremath{\mathsf{hz}}}
\newcommand\tp{\ensuremath{\mathsf{tp}}}
\newcommand\sev{\ensuremath{\mathsf{sv}}}
\newcommand\subhz{\ensuremath{\mathsf{subHz}}}
\newcommand\err{\ensuremath{\mathsf{err}}}
\newcommand\loss{\ensuremath{\mathsf{loss}}}
\newcommand\omission{\ensuremath{\mathsf{omission}}}
\newcommand\late{\ensuremath{\mathsf{late}}}
\newcommand\early{\ensuremath{\mathsf{early}}}
\newcommand\minor{\ensuremath{\mathsf{minor}}}
\newcommand\major{\ensuremath{\mathsf{major}}}
\newcommand\fatal{\ensuremath{\mathsf{fatal}}}
\newcommand\cat{\ensuremath{\mathsf{cat}}}
\newcommand\safMon{\ensuremath{\mathsf{safMon}}}
\newcommand\sm{\ensuremath{\mathsf{sm}}}
\newcommand\watchDog{\ensuremath{\mathsf{watchDog}}}
\newcommand\wdcp{\ensuremath{\mathsf{wd}}}
\newcommand\nProg{\ensuremath{\mathsf{nProg}}\xspace}
\newcommand\dProg{\ensuremath{\mathsf{2Prog}}\xspace}
\newcommand\voter{\ensuremath{\mathsf{vt}}}
\newcommand\VOTER{\ensuremath{\mathsf{VT}}}
\newcommand\fs{\ensuremath{\mathsf{fs}}}

This section introduces our domain-specific language, called \dsl, for enabling automated safety reasoning with safety patterns. 
Tables~\ref{tab:dsl1} and \ref{tab:dsl2} describe \dsl's main elements, \ie, key terms and predicates.
Table~\ref{tab:dsl1} describes the language used to specify functional and hardware architecture, and safety analysis, while Table~\ref{tab:dsl2} describes the predicates used to specify selected safety patterns. 
We illustrate \dsl\ by using the \acc example described in Section~\ref{sec:motivation}. 

%\dsl\ is available online at~\cite{?}.
\begin{table}[t]
  \begin{tabular}{p{2.3cm}p{12.6cm}}
    \toprule
    \multicolumn{2}{c}{\textbf{Functional, Hardware and Safety Analysis}}\\
    \midrule
    \textbf{Fact} & \textbf{Denotation}\\
    \midrule
    \cp(\id) & $\id$ is a function in the system.\\
    \midrule
    \subcp($\id_1$,$\id_2$) & $\id_1$ is a sub-function of the function $\id_2$.\\
    \midrule
    \ch(\id,$\id_1$,$\id_2$) & $\id$  is a logical channel connecting an output of the function $\id_1$ to an input of the function $\id_2$. Notice that it denotes a unidirectional connection.\\
    \midrule
    $\info(\id,\vec{\ch})$ & $\id$ is an information flow following the channels in $\vec{\ch}$.\\
    \midrule
    \hw(\id) & Function $\id$ is implemented as hardware, \eg, circuit connected to sensors.\\
    \midrule
    \sw(\id) & Function $\id$ is implemented as a software.\\
    \midrule
%    \ecu(\id) & $\id$ is an Electronic Computing Unit (ECU) that can run software functions.\\
%    \midrule
%    \can(\id) & $\id$ is a Controller Area Network (CAN) used to communicate between ECUs.\\
%    \midrule
%    \Dep(\id,$\id_h$) & Function $\id$ is deployed, \ie, executed, in the ECU $\id_h$ or alternatively, the logical channel $\id$ is deployed in the CAN $\id_h$ to establish the communication between $\id$'s functions.\\
%    \midrule

    \hz(\id,$\id_c$,\tp,\sev) & $\id$ is a hazard  associated with the function $\id_c$ is of type $\tp$, where $\tp \in \{\err,\loss,\omission,\late,\early\}$, and severity $\sev$, where $\sev \in \{\minor, \major, \fatal,\cat\}$. $\err$, $\loss$, $\omission$, $\late$, and $\early$ denote, respectively, erroneous, loss of function, omission, late and early types of hazards. $\minor, \major, \fatal,\cat$ denotes, respectively, minor, major, fatal and catastrophic severity levels.\\

    \midrule
    \subhz($\id_1$,$\id_2$) & $\id_1$ is a hazard causing hazard $\id_2$.\\
    \bottomrule
  \end{tabular}
  \caption{\dsl: a DSL for specifying functional, hardware and safety analysis.}
  \label{tab:dsl1}
\end{table}

\begin{table}
  \begin{tabular}{lp{10cm}}
    \toprule
    \multicolumn{2}{c}{\textbf{Safety Architectural Patterns}}\\
  \midrule
    \textbf{Fact} & \textbf{Denotation}\\
   \midrule

\dmoon(\id,$\id_c$,${I_c}$,$\id_{c\prime}$,$I_{\voter_{1}}$,$I_{\voter_{2}}$,$\voter$,\\$\voter_{out}$,$\id_{out}$) &
\vspace{-0.6cm}
$\id$ is a duplex redundancy associated with the function $\id_c$. ${I_c}$ is a channel from $\id_c$ that might convey a fault message. 
$\id_{c\prime}$ is a function possibly $\id_c$. 
$\voter$ is a voter that receives data from $\id_c$ and $\id_{c\prime}$ through channels $I_{\voter_{1}}$, and $I_{\voter_{2}}$, respectively. The result from $\voter$ is sent to $\id_{out}$ through channel $\voter_{out}$.  \\
  \midrule
  \tmoon(\id,$\id_c$,${I_c}$,$\id_{c\prime}$,$\id_{c\prime\prime}$,$I_{\voter_{1}}$,$I_{\voter_{2}}$,\\$I_{\voter_{3}}$,$\voter$,$\voter_{out}$,$\id_{out}$) &
\vspace{-0.6cm}
$\id$ is a triple modular redundancy associated with the function $\id_c$. ${I_c}$ is a channel from $\id_c$ that might convey a fault message. 
$\id_{c\prime}$ and $\id_{c\prime\prime}$ are functions possibly $\id_c$.
%$\id_{c\prime}$ and $\id_{c\prime\prime}$ are redundant functions of $\id_c$. 
$\voter$ is a voter that receives data from $\id_c$, $\id_{c\prime}$ and $\id_{c\prime\prime}$ through channels $I_{\voter_{1}}$, $I_{\voter_{2}}$, and $I_{\voter_{3}}$, respectively. The result from $\voter$ is sent to $\id_{out}$ through channel $\voter_{out}$.  \\
  \midrule
%  \dProg(\id,$\id_c$,$\vec{I_{\id_{c}}}$,$\id_c\prime$,$\vec{I_{\id_{c\prime}}}$,$\vec{I_{\voter_{1}}}$,$\vec{I_{\voter_{2}}}$,\\
  \dProg(\id,$\id_c$,$\vec{I_{\id_{c}}}$,$\vec{O_{\id_{c}}}$,$\id_c\prime$,$\vec{I_{\voter_{1}}}$,$\vec{I_{\voter_{2}}}$,\\
$\vec{\VOTER}$,$\vec{\VOTER_{out}}$,$\vec{\id_{out}}$) &
\vspace{-0.6cm} 
$\id$ is a 2-version programming associated with the function $id_c$ (a.k.a. version 1). $\id_{c\prime}$ (a.k.a. version 2) is an identical function of $\id_c$. 
The inputs to $id_c$ and the outputs from $id_c$ are sent through channels $\vec{I_{\id_{c}}}$ and $\vec{O_{\id_{c}}}$, respectively.
%The outputs from $id_c$ are sent through channels $\vec{O_{\id_{c}}}$. 
$\vec{\VOTER}$ is a list of voters that receive data from $\id_c$ and $\id_c\prime$ through channels $\vec{I_{\voter_{1}}}$ and $\vec{I_{\voter_{2}}}$, respectively. The results from $\vec{\VOTER}$ are sent to their respective functions $\vec{\VOTER_{out}}$ through channels $\vec{\id_{out}}$.\\

%$\id$ uses the list of input channels $\vec{I_{\id_{c}}}$ and $\vec{I_{\id_{c\prime}}}$ from $\id_c$ and $\id_c\prime$, respectively. $\vec{\VOTER}$ is a list of voters that receive inputs from $\id_c$ and $\id_c\prime$ via channels $\vec{I_{\voter_{1}}}$ and $\vec{I_{\voter_{2}}}$, respectively. The results from $\vec{\VOTER}$ are sent to their respective functions $\vec{\VOTER_{out}}$ via channels $\vec{\id_{out}}$.    \\
  \midrule
  \safMon(\id,$\id_c$,$\vec{I}$,$\vec{O}$,$\fs$,$\vec{I_\sm}$,$\vec{O_\sm}$,\sm) & 
$\id$ is a safety monitor associated with the function $\id_c$. It uses the list of input and output channels $\vec{I}$ and $\vec{O}$, respectively. The data of these channels are sent as input to $\sm$ through the list of channels $\vec{I_\sm}$ and $\vec{O_\sm}$. $\fs$ is a channel from $\sm$ to $\id_c$ which sends a fail-safe signal whenever some inconsistency is detected. \\
  \midrule
  \watchDog(\id,$\id_c$,$\fs$,${I_\wdcp}$,\wdcp) & 
$\id$ is a watchdog associated with the function $\id_c$. It receives liveness messages from $\id_c$ through channel ${I_\wdcp}$. $\fs$ is a channel from $\wdcp$ to $\id_c$ which sends a fail-safe signal whenever some inconsistency w.r.t the expected messages is detected. \\
  \bottomrule
  \end{tabular}
  \caption{\dsl: Language for Safety Architectural Patterns.}
  \label{tab:dsl2}
\end{table}

\begin{example}
\label{ex:dsl}
  The functional architecture depicted in Figure~\ref{fig:acc-func} is specified by the following atomic formulas, or facts, using the notation of the DLV prover~\cite{leone06tcl}:
\begin{verbatim}
   cp(acc).   cp(accm).   cp(ds).   cp(vs).   cp(bs).   cp(ps).
   subcp(accm,acc).   subcp(ds,acc).   subcp(vs,acc). ch(dsaccm,ds,accm).  
   ch(vsaccm,vs,accm).  ch(accmbs,accm,bs). ch(accmps,accm,ps). 
   if(if1,[vsaccm,accmbs]).  if(if2,[dsaccm,accmbs]). 
\end{verbatim}
The fact \texttt{ch(vsaccm,vs,accm)} denotes the logical communication between the \vs and the \accm. 
The information flow \texttt{if1} denotes data flows from \vs to \bs.
%Also, \red{data flows from the \vs to} , as specified by the information flow \texttt{if1}.
The facts below specify which functions are implemented as software, \eg, \accm, and which as hardware, \eg, \ds.
\begin{verbatim}
   sw(accm).   hw(ds).   hw(vs).   hw(ps).   hw(bs).
\end{verbatim}

\noindent Finally, the \acc hazards and their relations are specified by the following facts:

\begin{verbatim}
  hz(h1,acc,err,cat).  hz(h2,acc,loss,cat).  hz(h11,ds,err,cat).
  hz(h12,vs,err,cat).  hz(h13,accm,err,cat).
  subHz(h11,h1).  subHz(h12,h1).  subHz(h13,h1).
\end{verbatim}
For example, the hazard \accmerr (\texttt{h13}) is a sub-hazard of \accerr (\texttt{h1}).
\end{example}

\begin{wrapfigure}[10]{r}{0.45\textwidth}
%\vspace{-4mm}
%  \includegraphics[width=0.44\textwidth,natwidth=111,natheight=65]{figures/safmon.pdf}
  \includegraphics[width=0.44\textwidth]{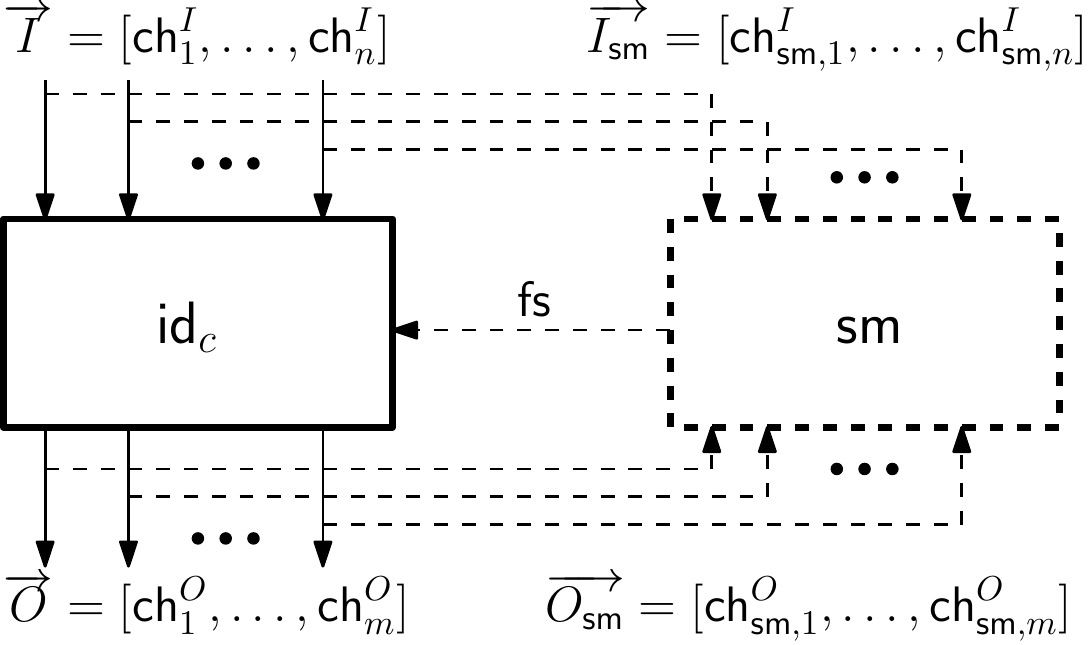}
  \vspace*{-2mm}
  \caption{Safety Monitor Pattern}
  \label{fig:safmon}
\end{wrapfigure}

Due to space limitations, we illustrate only the \safMon\ pattern. 
The remaining patterns follow a similar reasoning. We refer the reader to~\cite{preschern13plop,valdivia14dasc} for detailed description of these patterns. 

The \safMon\ pattern is depicted by all dashed elements in Figure~\ref{fig:safmon} including channels. This \safMon\ is associated to the function $\id_c$ and is used to detect whether $\id_c$ is computing erroneous values. 
To this end, it takes the values of $\id_c$'s inputs ($\vec{I}$) and outputs ($\vec{O}$) to the function $\sm$ through the channels $\vec{I_\sm}$ and $\vec{O_\sm}$. The channel $\fs$ connecting $\sm$ with $\id_c$ is used to send fail-safe commands whenever abnormal input-output relations are detected by \sm. 
%Notice that the channel and component identifiers are assumed to be disjoint. 

In \dsl, one identifies a \safMon\ by specifying the fact \safMon(\id,$\id_c$,$\vec{I}$,$\vec{O}$,$\fs$,$\vec{I_\sm}$,$\vec{O_\sm}$,\sm), containing all the information related to the safety monitor as described above.

\section{Safety Reasoning using DLV}
\label{sec:automation}
%!TEX root = iclp20.tex

\newcommand\ctl{\ensuremath{\mathsf{ctl}}}
\newcommand\nctl{\ensuremath{\mathsf{nctl}}}

% \subsection{Reasoning Principles}
%One of the main goals of safety engineers during the definition of a system architecture is to place suitable safety patterns, so that the identified hazards can be controlled provided the patterns are implemented as expected, \eg, are bug free. This section demonstrates how much of this safety reasoning can be automated.

One of the main goals of safety engineers during the definition of a system architecture is to place suitable safety patterns so that the identified hazards can be controlled.
This section demonstrates how much of this safety reasoning can be automated.

To this end, we introduce two new facts used to denote when a hazard is controlled or not:
\begin{itemize}
  \item $\ctl(\id_H,\id_c,\tp,\sev)$ and $\nctl(\id_H,\id_c,\tp,\sev)$ denote that the hazard $\id_H$ of type $\tp$, severity $\sev$ and associated with the function $\id_c$ can be, respectively, controlled and not controlled.
\end{itemize}

Before we specify controlled and not controlled hazards, we need to distinguish two types of hazards: \emph{basic} hazards and \emph{derived} hazards. A hazard is classified as \emph{basic} if it does not have any sub-hazards, and \emph{derived} otherwise. The following DLV rules specify this:
\begin{verbatim}
  basic(H,CP,TP,SV) :- hz(H,CP,TP,SV), not has_subHz(H).
  has_subHz(H) :- subHz(SH,H).
  derived(H,CP,TP,SV) :- hz(H,CP,TP,SV), has_subHz(H).
\end{verbatim}

We now use the closed-world semantics of DLV to specify controllability. A basic hazard is not controlled if there is no rule explicitly supporting its controllability, as specified by the rule:

\begin{verbatim}
  nctl(H,CP,TP,SV) :- basic(H,CP,TP,SV), not ctl(H,CP,TP,SV) .
\end{verbatim}

A derived hazard is not controlled if any one of its sub-hazards is not controlled as specified by the following rules:

\begin{verbatim}
  nctl(H,CP,TP,SV) :- hz(H,CP,TP,SV), derived(H,CP,TP,SV), 
                      hasNCTLSubHz(H,CP,TP,SV).
  hasNCTLSubHz(H,CP,TP,SV) :- hz(H,CP,TP,SV), subHz(SH,H),
                              nctl(SH,SCP,STP,SSV).
\end{verbatim}

\begin{example}
 Consider the hazards and sub-hazards relations in Example~\ref{ex:dsl}. The hazards \texttt{hz(h1,acc,err,cat)} can be controlled if its three sub-hazards, \texttt{h11}, \texttt{h12} and \texttt{h13}, can be controlled.
\end{example}

Safety patterns are commonly used to control hazards by, \eg, adding redundancy to the system. Given our language \dsl, the reasoning principles used to do so can be easily captured by DLV rules.
We list some reasoning principles for some of the patterns: 

%\noindent
\paragraph{WatchDog Pattern.} The following rule specifies that watch dog can be used to control hazard of type loss of function (\texttt{loss}).
\begin{verbatim}
  ctl(ID,CP,loss,SV) :- hz(ID,CP,loss,SV), watchDog(_,CP,_,_,_).  
\end{verbatim} 

%\noindent
\paragraph{Safety Monitor Pattern.} The following rules specify intuitively that a hazard associated to a function \texttt{CP} of type erroneous can be controlled if a safety monitor is associated to \texttt{CP} provided \texttt{not inpNotCovSF(ID2)} and \texttt{not outNotCovSF(ID2)}: there are no input logical channels, \ie, channels incoming to \texttt{CP} specified by \texttt{ch(CH,\_,CP)}, not taken as input to the safety monitor, nor output channels \ie, channels outgoing from \texttt{CP} specified by \texttt{ch(CH,CP,\_)}.
The predicate \texttt{\#member}, \eg, \texttt{\#member(CH,ICHs)} specifies that \texttt{CH} is a member of list \texttt{ICHs}.
You can safely ignore the fact \texttt{isexploration} which is only used for the automation as described in Section~\ref{sec:recommendation}.
\begin{verbatim}
  ctl(ID,CP,err,SV) :- hz(ID,CP,err,SV), safMon(ID2,CP,_,_,_,_,_,_),
                       not inpNotCovSF(ID2), not outNotCovSF(ID2).
  inpNotCovSF(ID2) :-  safMon(ID2,CP,ICHs,_,FS,_,_,_), ch(CH,_,CP),
                       CH != FS, not #member(CH,ICHs), not isexploration.
  outNotCovSF(ID2) :- safMon(ID2,CP,_,OCHs,_,_,_,_), ch(CH,CP,_),
                      not #member(CH,OCHs), not #member(CH,MIN),
                      not #member(CH,MOUT), not isexploration.
\end{verbatim}

%\noindent
\paragraph{2-version programming.} This pattern is used to improve safety by adding software redundancy. Hence, it can only be associated with functions implemented as software as specified by the rule:
\begin{verbatim}
  ctl(ID,CP,err,SV) :- hz(ID,CP,err,SV), 2Prog(ID2,CP,_,_,_,_,_,_),
        sw(CP), not inpNotCovNP(ID2).
\end{verbatim}

\noindent Here \texttt{inpNotCovNP} is similar to \texttt{inpNotCovSF} explained above.

%Here \texttt{inpNotCovNP} and \texttt{outNotCovNP} are similar to \texttt{inpNotCovSF} and \texttt{outNotCovSF} above.
%\noindent
\paragraph{HDR.} The \dmoon and \tmoon Voter patterns can used for two different safety reasons: (1) to improve safety by hardware redundancy or (2) to improve safety by path redundancy. These are specified by the following rules, where \texttt{omission} is a type of error:
\begin{verbatim}
  ctl(ID,CP,err,SV) :- hz(ID,CP,err,SV), 
          hdr(ID3,_,_,_,_,_,VOTERCP,_,_), ch(_,CP,VOTERCP).
  ctl(ID,CP,omission,SV) :- hz(ID,CP,omission,SV),cp(CP),cp(CP1),cp(CP2),
         ch(CHOUT,CP1,_), ch(CHIN,_,CP2), ch(CH,CP,_),if(IF,PATH),
         before(CH,CHIN,IF), before(CHOUT,CH,IF),
         hdr(IDPAT,CP1,_,CP2,_,_,_,_,_).
\end{verbatim}
The second rule requires further explanation. 
The fact \texttt{before}, \eg, \texttt{before(CH,CHIN,IF)} specifies that \texttt{CH} appears before \texttt{CHIN} in the path \texttt{PATH}. 
The rule itself specifies that if there is an \texttt{IF} such that there is a hazard of type \texttt{omission} associated to a component \texttt{CP} in the information path \texttt{PATH}, then placing a $\dmoon$ on a functions \texttt{CP1} and \texttt{CP2} before and after \texttt{CP} in the path can control an omission hazard. 
Intuitively, this is because Voters places in this way can detect when safety critical messages are lost during transmission due to the omission of \texttt{CP}.\\

\noindent\textbf{Remark:}
This paper specifically focuses on architectural principles. 
We focus on the architecture components and how such components interact with other through channels.
Encoding other reasoning principles like, \eg, the actual behavior of such components, are left to future work.

\section{Automated Pattern Recommendation}
\label{sec:recommendation}
%!TEX root = iclp20.tex

\newcommand\explore{\ensuremath{\mathsf{explore}}}
\newcommand\N{\ensuremath{\mathsf{N}}}
\newcommand\Pat{\ensuremath{\mathsf{Pat}}}

This section builds on the principles specified to automate the recommendation of safety pattern.
Our machinery enables a safety engineer to understand which options of patterns he can use to control hazards and decide which one is more suitable given factors, such as costs and hardware availability.

The recommendation machinery uses ASP/DLV semantics to enumerate design options by attempting to place safety patterns wherever they are applicable.
In this way, each answer of our DLV specification corresponds to a recommended architecture. Some recommended architectures may be better than others, \eg, controlling more hazards.
From all obtained answers, the system can recommend to the safety engineer only the best architectures, \ie, the ones that control the most number of hazards.  

The recommendation system is activated by using facts of the form. 
\begin{itemize}
  \item $\explore(\N,\Pat)$ denoting that the system shall recommend the placement of at most $\N$ patterns of type $\Pat$, where $\Pat$ is one of patterns described in Table~\ref{tab:dsl2}. 
\end{itemize}
%\explore's arguments enable the control of the search space used by the system.
For example, if \explore(1,\safMon), the system attempts to add at most one additional safety monitor to a given architecture.
Multiple such facts can be used to recommend different patterns at the same time.
As a result, safety engineers can configure the pattern recommendation machinery to search for particular safety patterns that can control identified hazards.
 
%We have implemented rules for recommending all types of patterns shown in Table~\ref{tab:dsl2}. Due to space restrictions, we describe only some of them used for recommending \safMon\ and \dmoon. 
We have implemented rules for recommending the patterns shown in Table~\ref{tab:dsl2}. Due to space restrictions, we describe only some of them used for recommending \safMon, \tmoon, and \dmoon. 

%The following DLV rule specifies the enumeration of placement or not of a \safMon, denoted by \texttt{nsafMon}, associated with the function \texttt{CP} that is furthermore associated with a basic or not controlled hazard \texttt{ID}:
The following DLV rule specifies the enumeration of placement or not of a \safMon (\texttt{nsafMon}), associated with the function \texttt{CP} that is furthermore associated with a basic or not controlled hazard \texttt{ID}:

\begin{verbatim}
 safMon(nuSafMon,CP,allInputs,allOutputs,nuSC,numin,numout,numcp) v 
 nsafMon(nuSafMon,CP,allInputs,allOutputs,nuSC,numin,numout,numcp) 
 :- cp(CP),hz(ID,CP,err,SV),basicOrNCTL(ID,CP,err,SV),explore(N,safMon).
\end{verbatim}
We assume here that the constants starting with \texttt{nu} are fresh, \ie, do not appear in the given architecture, thus used only for recommended safety patterns. Since it is enough to know to which function a safety monitor is associated to, we do not need to enumerate all the inputs and outputs of \texttt{CP}, but rather simply denote \texttt{CP}'s inputs and outputs using, respectively, the fresh constants \texttt{allInputs} and \texttt{allOutputs}.

The rule above will attempt to place a safety monitor in any applicable location of the architecture. 
The following clause limits the number of safety monitors that can be recommended to be at most \texttt{N}. Here \texttt{\#count} is a DLV aggregate predicate returning the size of a symbolic set defined by its argument.
\begin{verbatim}
  :- #count{CP : safMon(nuSafMon,CP,_,_,_,_,_,_)} > N, explore(N,safMon).
\end{verbatim}

%The rule above constraints the number of safety monitors by \texttt{N}.  \texttt{CP} to which a new safety monitors are associated with. 
%The rule above returns the number of \texttt{CP} to which a new safety monitors are associated with. 
%In this rule, above it returns the number of \texttt{CP} to which a new safety monitors are associated with.

Notice that whenever a pattern is recommended, the controllability reasoning described in Section~\ref{sec:automation} applies to infer which hazards are controlled by this pattern and which are not.

The reasoning principles described in Section~\ref{sec:automation} can be used to further constraint the number of recommendations. For example, a $\tmoon$ used for hardware redundancy shall only be associated with components that are not software components as specified by the following rule:
\begin{verbatim}
tmr(nuTMR,CP1,CH1,nucp2,nucp3,nuchm1,nuchm2,nuchm3,nuvtcp,nucho,nucpo) v 
ntmr(nuTMR,CP1,CH1,nucp2,nucp3,nuchm1,nuchm2,nuchm3,nuvtcp,nucho,nucpo)
:- cp(CP1),not sw(CP1),hz(HZ0,CP1,err,SV), ch(CH1,CP1,_), explore(N,tmr).
\end{verbatim}

The next example illustrates the power of our language to specify pattern recommendation. It specifies conditions for recommending \dmoon\ patterns to achieve path redundancy. 
\begin{verbatim}
  hdr(nuHDR,CP1,CH1,CP2,nuchm1,nuchm2,nuvtcp,nucho,CPO)
  v nhdr(nuHDR,CP1,CH1,CP2,nuchm1,nuchm2,nuvtcp,nucho,CPO)
  :- hz(ID,CP,omission,SV), cp(CP), cp(CP1), cp(CP2), CP1 != CP,
     CP1 != CP2, CP1 != CPO, CP2 != CPO, ch(CHOUT,CP1,_), ch(CHIN,_,CP2),
     ch(CH,CP,_), ch(CH1,_,CPO), if(IF,PATH), before(CHOUT,CHIN,IF),
     before(CHOUT,CH,IF), before(CHIN,CH1,IF), explore(N,hdr).
\end{verbatim}
We search for functions \texttt{CP0}, \texttt{CP1} and \texttt{CP2} and a channel \texttt{CH1} where to place the \dmoon. The goal is to control a hazard associated with function \texttt{CP} by increasing path redundancy. 
To this end, \texttt{CP1} needs to appear before \texttt{CP} in an information flow \texttt{PATH} that uses these functions. 
\texttt{CP2} may either be equal to \texttt{CP} or located after \texttt{CP} in such a \texttt{PATH}.
%This means that \texttt{CP1} and \texttt{CP2} have to appear before and after \texttt{CP} in an information flow \texttt{PATH} that uses these functions. 
Thus, \dmoon can, in principle, detect when messages are omitted by \texttt{CP}. 
Whenever this happens, the \dmoon shall send a message to the function \texttt{CPO} used only later in the information flow \texttt{PATH}.

A constraint similar to the one for \safMon, constraints the number of \tmoon and \dmoon to be searched for. These constraints are omitted here. \\

\section{Case Studies}
\label{sec:examples}
%!TEX root = iclp20.tex

This section illustrates the results of our automated safety reasoning for two case studies, namely Adaptive Cruise Control (ACC) and Battery Management System (BMS).
We illustrate our results by depicting how the architectures of both \acc and \bms would appear on a layout when our machinery is used.
The safety patterns suggested by our machinery are depicted as dark gray boxes, and the channels related (inputs or outputs) to such patterns are depicted as dashed arrows.

\paragraph{Adaptive Cruise Control (ACC). } We identified an erroneous (\accerr) and a loss (\accloss) hazard on \acc, as described in Section~\ref{sec:motivation}.
The erroneous hazard (\accerr) is broken down into three sub-hazards, namely erroneous \ds (\derr), erroneous \vs (\verr), and erroneous \accm (\accmerr).

\begin{figure}[h]
  \centering							
  \includegraphics[width=0.65\textwidth]{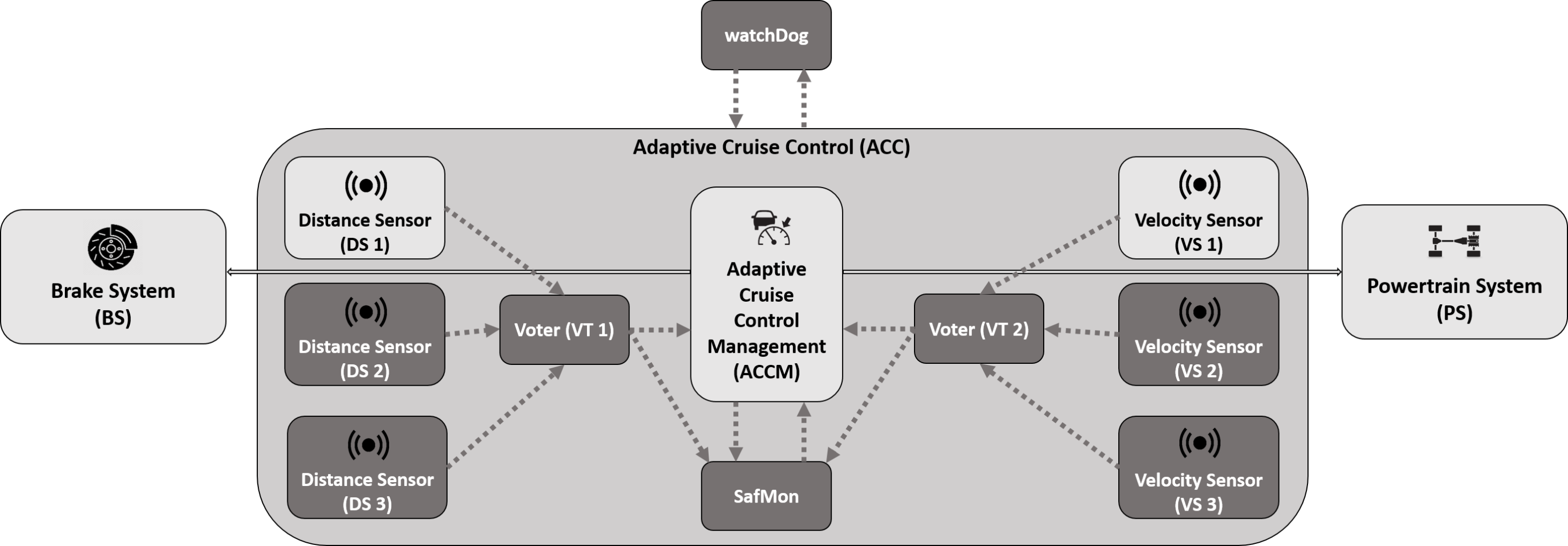}
  \caption{ACC Functional Architecture with \safMon, \tmoon and \watch}
  \label{fig:acc-func-machinery}
\end{figure}

We run our recommendation machinery to automatically identify what safety patterns could be used to control the identified hazards.
%\accerr, \accloss, \derr, \verr, and \accmerr.
Our machinery yielded \emph{five complete solutions} (\ie, architectures) for controlling these hazards.
For the sake of space, we only show one of those solutions. 
The architecture of the chosen solution is depicted in Figure~\ref{fig:acc-func-machinery}.
%The subset of the model for this solution containing only the predicates on suggested safety patterns and controllability is shown below.
The subset of our DLV specification for this solution is shown below. 
It contains the predicates for the recommended safety patterns and controllability.

\begin{verbatim}
{safMon(nuSafMon,accm,allInputs,allOutputs,nuSC,numin,numout,numcp),
tmr(nuTMR,ds,dsaccm,nucp2,nucp3,nuchm1,nuchm2,nuchm3,nuvtcp,nucho,nucpo),
tmr(nuTMR,vs,vsaccm,nucp2,nucp3,nuchm1,nuchm2,nuchm3,nuvtcp,nucho,nucpo),
watchDog(nuWD,acc,nuscwd,nulvwd,nuwd), ctl(["hz",accLs],acc,loss,cat),
ctl(["hz",ds],ds,err,cat), ctl(["hz",vs],vs,err,cat),
ctl(["hz",accm],accm,err,cat), ctl(["hz",accEr],acc,err,cat)}
\end{verbatim}

Our machinery recommended to use three safety patterns, \ie, \safMon, \tmoon, and \watchDog, to control the identified hazards.
The main difference w.r.t. the other solutions (omitted here) is \dProg instead of \safMon.
%The main difference w.r.t. the other solutions (omitted here) is the suggestion of \nProg.
To control the sub-hazards \derr and \verr, our machinery recommended to use \tmoon on \ds and \vs, respectively.
The goal is to improve safety by hardware (\ie, \ds and \vs) redundancy.
The remaining sub-hazard \accmerr can be controlled by placing a \safMon\ on \accm. 
%A \safMon is recommended to monitor the behavior of \accm so that \accmerr is controlled.
%A \nProg is recommended to monitor the behavior of \accm so that \accmerr is controlled.
The hazard \accerr is then controlled by using both \tmoon and \safMon.
%As a result, \accerr can be controlled.
Finally, our machinery recommended to use a \watchDog\ on \acc to control the loss hazard \accloss.

\paragraph{Battery Management System (BMS).} We identified an erroneous (\cierr) hazard on \ci, as described in Section~\ref{sec:motivation}.
This erroneous hazard (\cierr) is broken down into three sub-hazards, namely erroneous \bms (\bmserr), erroneous \can\ (\canerr), and omission \fw (\fwerr). 
Typically, hazards on \can\ buses can be controlled by replacement only. 
Hence, we assume that \canerr has already been controlled.

%do not consider a safety pattern for controlling \canerr.

Our recommendation machinery yielded \emph{four complete solutions} (\ie, architectures) to control \cierr, \bmserr, and \fwerr.
For the sake of space, we only show two of those solutions. 
The architecture of the chosen solutions are depicted in Figure~\ref{fig:acc-func-machinery}.
The DLV specification for those solutions is similar to the one presented in the ACC case study.
% Section~\ref{subsec:acc}.

\begin{figure}[ht]
\begin{subfigure}{.48\textwidth}
  \centering
  % include first image
  \includegraphics[width=.65\linewidth]{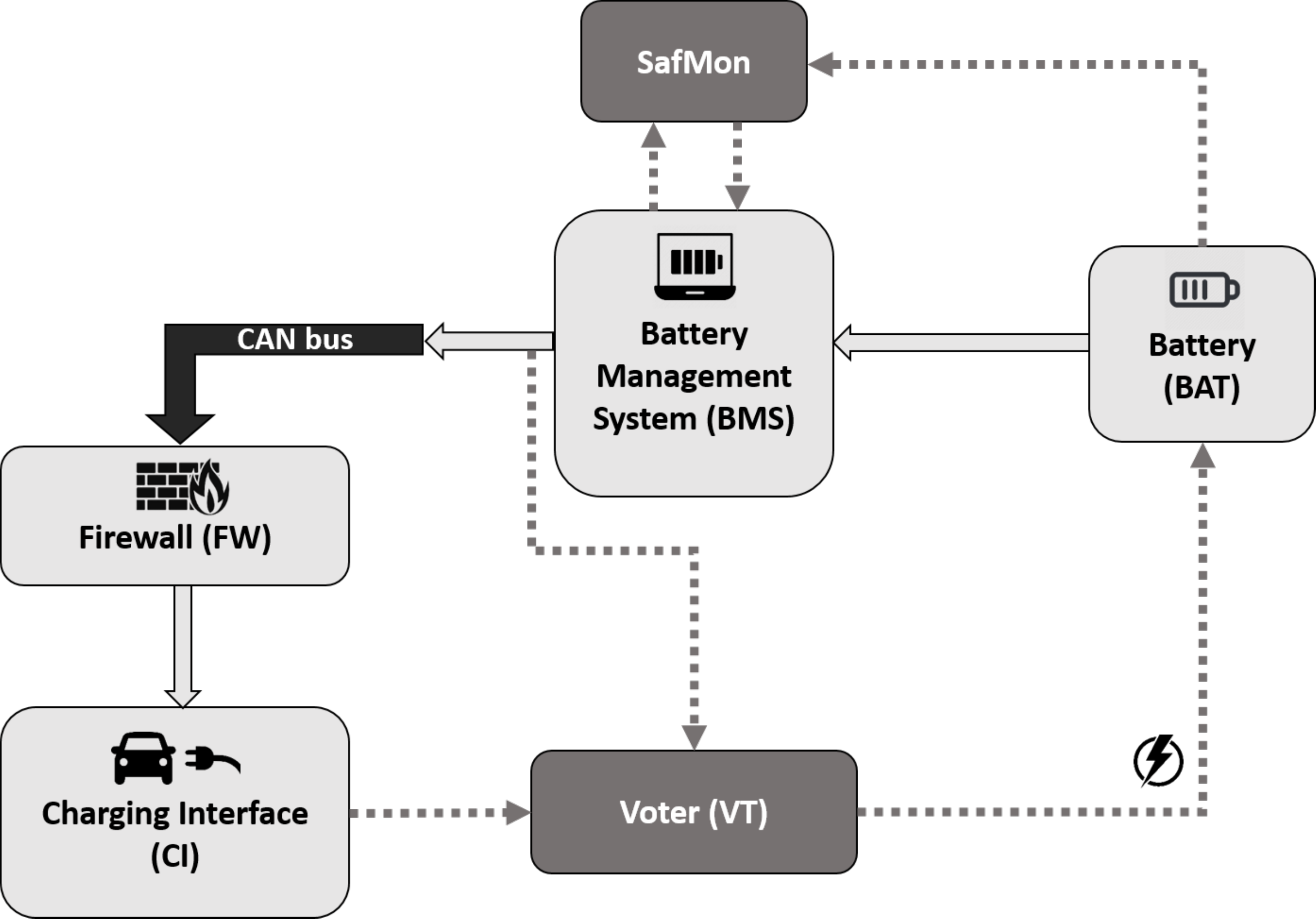}    
  \caption{Path redundancy for \bms and \ci}
  \label{fig:bms-func-machinery-1}
\end{subfigure}
\begin{subfigure}{.48\textwidth}
  \centering
  % include second image
%  \includegraphics[width=.65\linewidth,natwidth=298,natheight=208]{figures/battery-safmon-dmoon-option2.pdf}
  \includegraphics[width=.65\linewidth]{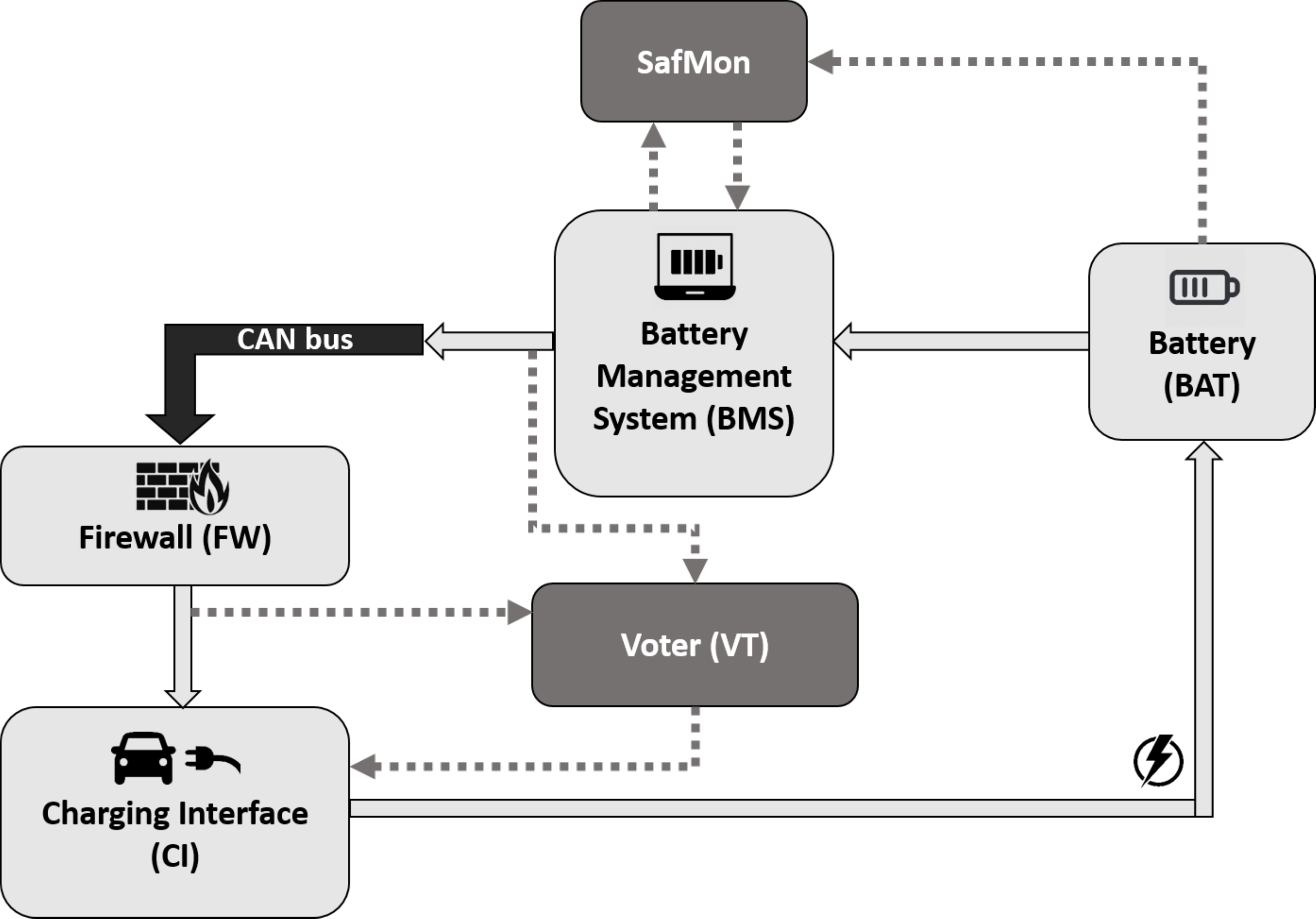}  
  \caption{Path redundancy for \bms and \fw}
  \label{fig:bms-func-machinery-2}
\end{subfigure}
\caption{Battery Management System Functional Architecture with \safMon\ and \dmoon}
\label{fig:bms-func-machinery}
\end{figure}

Our machinery recommended to use two safety patterns, \ie, \safMon\ and \dmoon, to control the identified hazards.
On both solutions, a \safMon\ is placed together with \bms to control \bmserr.
For the \acc example, \tmoon is placed to improve safety by hardware redundancy.
%For the \acc example, we recommended to use \tmoon to improve safety by hardware redundancy.
Here, \dmoon is placed to improve safety by path redundancy. 
The \dmoon solutions are depicted in Figures~\ref{fig:bms-func-machinery-1} and \ref{fig:bms-func-machinery-2} control \fwerr.
They differ w.r.t which functions are composing the \dmoon.
Figure~\ref{fig:bms-func-machinery-1} illustrates that \bms and \ci sent redundant inputs to \voter\ so that \bat has a higher chance of getting the expected input.
That is, if \ci does not send the input to \bat due to, \eg, an omission from \fw, \bat receives the expected input from \bms through \voter. 
Similarly, Figure~\ref{fig:bms-func-machinery-2} illustrates that \bms and \fw sent redundant inputs to \voter\ with \ci as destination. Consequently, \bat should have a higher chance of getting the expected input from \ci.

\section{Related Work}
\label{sec:related}
%!TEX root = iclp20.tex

\paragraph{Failure Rates Computations.}
An important analysis for safety is the computation of failure rates of the system and its sub-systems as it is a requirement for safety-critical systems to have (very) low failure rates.
The automation of this computation has been subject of some previous work~\cite{helle,ft+}. In particular, for a given architecture and given sub-system fault rates, the failure rate of the system is computed.  
%The automation of this computation has been subject of some previous work~\cite{helle,ft+,weber}. In particular, for a given architecture and given sub-system fault rates, the failure rate of the system is computed. 
Our work on reasoning with safety patterns complements the work above as we consider the design of the architecture itself, which is part of the input used by the work above.
 
\vspace*{-0.4cm}

\paragraph{Safety Case Models.}
%\begin{comment} 
%\end{comment}
GSN~\cite{gsn11standard} is a model for specifying safety cases. Safety cases are tree-like structures containing different types of nodes denoting, \eg, Goals, Strategies, Contexts, Assumptions of a safety case. As the exact meaning of each node is specified textually (inside the node), models written in GSN enables little automation. There are, however, work that provide more structure to GSN models and others providing means for some automation~\cite{carlan19wosocer}. We describe some approaches below.
\cite{gleirscher17patterns} proposes patterns encoding typical safety reasoning principles, such as those using FTA, FMEA, STPA.
While these reasoning patterns provide some structure to GSN models, they suffer from the same automation limitations of GSN described above. 
On the one hand, our work complements this work by specifying reasoning principles based on safety patterns, which was not considered in~\cite{gleirscher17patterns}. 
On the other hand, we believe that it is possible to encode some of the reasoning principles described in \cite{gleirscher17patterns} and consider not only safety reasoning with patterns but the other types of reasoning described in~\cite{gleirscher17patterns}.  
\cite{duan17survey,durrwang17safecomp} propose automated quantitative evaluation methods for GSN models that associated to Goal nodes with values for belief, disbelief and uncertainty. 
It is not clear from this work how these values are related to the quality of safety argument.
We believe that the encoding of our reasoning principles can profit from this work to make the relation between the quality of the safety argument and the belief values more explicit. 

\vspace*{-0.4cm}

\paragraph{Safety Reasoning using Logic Programming.}
Logic programming has been used in the past for safety reasoning. For example,~\cite{Gmez2014AssuringSI} provides decision support for air traffic control systems by specifying landing criteria in complex landing situations by using Defeasible Logic Programming
(DeLP). 
~\cite{deLP} outlines a method for safety assessment
of medical devices also based DeLP. An interesting work is presented in~\cite{gallina} on the formalization of automotive standard requirements~\cite{iso26262} to enable automatic reasoning about compliance with the standard.
We take a similar approach to these works as we also use logic programming and engine to support safety engineers in the designing system architecture. 
However, we do not consider here reasoning with uncertain and incomplete knowledge as in the work above using DeLP. 
As described above, we are considering extending the type of safety reasoning encoded to also include uncertainty~\cite{duan17survey,durrwang17safecomp}.
DeLP is a method we could consider for modeling such arguments.
% Different from our work, the focus here is on the development life-cycle of software from the project management perspective.

\section{Conclusion}
\label{sec:conc}
%!TEX root = iclp20.tex
This paper establishes the first steps towards automated safety (and security) for embedded systems. 
We propose a domain-specific language, called \dsl, for safety reasoning on the architectural level using safety patterns. We encode typical safety reasoning principles as disjunctive logic programs, using these specification for increasing automated reasoning, namely, on determining controllability and recommending patterns.

We are currently investigating a number of future directions. 
We are considering other types of safety reasoning, \eg, reasoning with uncertainty. 
Further, as illustrated by the \bms case study, there are a number of co-analysis reasoning deriving from the use safety and security patterns. 
It seems possible to build on the grounds established by this paper to carry out such reasoning in an automated fashion.
%, \eg, by extending appropriately our DSL and reasoning principles.

The increased automation provided by our methods seems to support incremental methods for safety (and in the future security). 
It is possible to identify, \eg, which hazards are no longer controlled whenever there is an incremental change to the system. 
We are currently investigating how to improve the proposed automated reasoning for this purpose.
Finally, we plan to integrate our machinery into the Model-Based Engineering Tool AutoFOCUS3~\cite{af3}.
The goal is to enable safety engineers to use our automated reasoning with models written in AutoFOCUS3. 
This will also enable the use of automated methods for building safety cases modeled in GSN~\cite{carlan19wosocer}.

%\section*{Acknowledgment}
\paragraph{Acknowledgment.}
This project has received funding from the European Union's Horizon 2020 research and innovation programme under grant agreement No 830892. Nigam is partially supported by CNPq grant 303909/2018-8.

\bibliographystyle{eptcs}
\bibliography{generic}
\end{document}